# A Tentative Role for FOXP2 in the Evolution of Dual Processing Modes and Generative Abilities

**Courtney Chrusch (courtney_chrusch@hotmail.com) and Liane Gabora (liane.gabora@ubc.ca)**
Department of Psychology, University of British Columbia
Arts Building, 333 University Way, Kelowna BC, V1V 1V7, CANADA

## Abstract

It has been suggested that the origins of cognitive modernity in the Middle/Upper Paleolithic following the appearance of anatomically modern humans was due to the onset of dual processing or *contextual focus (CF),* the ability to shift between different modes of thought: an explicit mode conducive to logical problem solving, and an implicit mode conducive to free-association and breaking out of a rut. Mathematical and computational models of CF supported this hypothesis, showing that CF is conducive to making creative connections by placing concepts in new contexts. This paper proposes that CF was made possible by mutation of the FOXP2 gene in the Paleolithic. FOXP2, once thought to be the "language gene", turned out not to be uniquely associated with language. In its modern form FOXP2 enabled fine-tuning of the neurological mechanisms underlying the capacity to shift between processing modes by varying the size of the activated region of memory.

**Keywords:** Associative thought; Contextual focus; Creativity; Divergent thought; Dual process; FOXP2; Human evolution; Language evolution; Paleolithic; Neural basis of language

## Introduction

The realization that FOXP2 was not the "language gene" generated much-needed sober discussion about the simplicity of single-gene explanations of complex traits. However, if one finds a relatively simple explanation that is consistent with a wealth of data from multiple disciplines, it is parsimonious to start with that and see if it needs to be modified or elaborated. We suggest that while FOXP2 is not the 'language gene' it may have a broad but well-defined impact on cognition. This paper synthesizes genetic, neurological, cognitive, and anthropological research into an integrated account of the cognitive changes underlying behavioral modernity, including language, came about. Specifically, we propose that it enabled the onset of *contextual focus*—the ability to shift between explicit / analytic and implicit / associative modes of thought—which paved the way for not just language, but a range of cognitive abilities considered by anthropologists to be diagnostic of behavioral modernity.

First we place the discussion in its historical context by reviewing both the studies that implicated the FOXP2 gene in the evolutionary origins of language, and the evidence that caused this connection to come into question. Second, we review anthropological and archaeological evidence that the coming into prominence of creative and cognitive abilities (including but not limited to those that involve language) coincides with the evolutionary origins of FOXP2. Third, we review literature on dual processing theories and generative abilities. Finally, we synthesize these literatures in a new explanation of the role of FOXP2 in language specifically and cognitive development more generally.

## The Search for a Genetic Basis for Language

FOXP2 is a transcription gene on chromosome 7 (Reimers-Kipping et al., 2011). It regulates the activity of other genes that play a role in the development and function of the brain (Fisher & Ridley, 2013; Kovas & Plomin, 2006). It is associated with proper functioning of the motor cortex, the striatum, and the cerebellum, which controls fine motor skills (Liegeois, Morgan, Connelly, & Vargha-Khadem, 2011). It was proposed that the FOXP2 gene plays an important role in language acquisition when a mutation in this gene was associated with language impairment in a British Family known as the KE Family (Lai, Fisher, Hurst, Vargha-Khadem, & Monaco, 2001). Approximately half the members of this family, across several generations, exhibited verbal dyspraxia and severe difficulty in controlling the orofacial muscles required for speech articulation (Lalmansingh, Karmakar, Jin, & Nagaich, 2012). The family was diagnosed with Specific Language Impairment: a significant deficit in language development that exists despite adequate educational opportunity and normal nonverbal intelligence (Lai *et al.,* 2001; Morgan, 2013). Those affected by SLI show deficits to the articulation of speech sounds, verbal expression, comprehension of speech, and have trouble controlling the movement and sequencing of orofacial muscles resulting in deficits in fluent speech. Transcript sequencing showed that a mutation of FOXP2 widespread in the KE family results in language deficiencies (Lai *et al.,* 2001). FOXP2 thus became prematurely referred to as the "language gene" (Bickerton, 2007).

However, from the start there was question concerning its involvement in language because as a transcription gene it only has an *indirect* effect on neural structure or behavior (Bickerton, 2007; Reimers-Kipping et al., 2011). Moreover, although FOXP2 is involved in motor control and learning, there ae indications that it has a role in other abilities that do not involve language (Kurt, Fisher, & Ehret, 2012). Indeed, the KE Family demonstrated minor non-verbal disabilities such as lack of cognitive fluidity, and below-average IQ (Lai et al., 2001). It was concluded that the neurological basis of deficits associated with FOXP2 lie in the structural and functional abnormalities of cortico-striatal and cortico-cerebellar circuitries of the brain, which are important for learning, memory, and motor control, not language exclusively.

Language is likely a polygenic trait (Chabris et al., 2012; see Kovas & Plomin, 2006). It is difficult to identify all the genes that contribute to an ability as complex as language; this is the problem of "missing heritability" (Manolio et al., 2009). Moreover, a gene such as FOXP2 may be pleiotropic in its effects, *i.e.,* it may affect multiple traits (Kovas & Plomin, 2006). Nevertheless, it is well established that a small change or perturbation can percolate through a system resulting in massive, large-scale changes, a phenomenon known as self-organized criticality (Bak, Tang, & Wiesenfeld, 1987). Self-organized criticality plays an important role in gene expression, particularly in the case of regulatory genes (Kauffman, 1993). This makes it worth at least considering the possibility that FOXP2 plays a role in cognition that extends beyond its role in language and yet is nonetheless clear-cut and identifiable.

## Imitation and the Evolution of Language

Evidence that FOXP2 is actively transcribed in brain areas where mirror neurons are present led to the hypothesis that FOXP2 made language possible through its effect on the capacity to imitate (Corballis, 2004a). Two simple forms of imitation are: (1) *onomatopoeia,* in which a speaker refers to something by making a sound that phonetically resembles it (e.g., referring to the sound made by a frog as "ribbit"), and (2) *interjections,* which may have begun with sounds such as the smacking of the mouth and lips during hunger. It has long been thought that onomatopoeia and interjections paved the way for more complex vocalizations (Chomsky, 1975; Farrar, 1866). The idea that imitation precipitated language was strengthened by findings that mirror neurons play a key role in language development (Arbib, 2011). However, the role of mirror neurons is controversial and the evidence for transcription of FOXP2 in areas where they are present is disputed. Moreover, in order to imitate the language of others there must already be others who are using language. Therefore, if FOXP2 played a role in the origin of language this is not fully accounted for by the mediating role of imitation. Moreover, the hypothesis that FOXP2 affects language by way of its effect on imitation does not explain the existence of defects associated with FOXP2 mentioned earlier that involve neither language nor imitation. Finally, imitation, but not grammatical, syntactically rich language, is seen in other species. In short, FOXP2 does not appear to affect language solely through its effect on imitation.

## The Cultural Explosion of the Middle-Upper Paleolithic

FOXP2 underwent accelerated evolution approximately 200,000 years ago (Corballis, 2004b; Lai et al., 2001). Within the last 200,000 years it underwent at least two human-specific mutations (Morgan, 2013; Lai *et al*., 2001). This corresponds to the appearance of anatomically modern humans in the fossil record between 200,000 and 100,000 years ago (Lai et al., 2001). It has been proposed that the appearance of anatomically modern humans was due to amino acid substitutions that differentiate the human FOXP2 gene from that of chimpanzees (Enard et al., 2002; Zhang et al., 2002). At least one FOXP2 mutation occurred within the last 100,000 years (Lai *et al*., 2001). This aligns FOXP2 change with the onset of behavioral modernity including complex language. Anatomical modernity preceded behavioral modernity; Leakey (1984) writes of anatomically modern human populations in the Middle East with little in the way of culture, and concludes, "The link between anatomy and behavior therefore seems to break" (p. 95).

The origin of complex language is widely associated with what Mithen (1998) refers to as the 'big bang' of human culture. The European archaeological record provides extensive evidence of an unparalleled cultural transition occurred between 60,000 and 30,000 years ago, at the onset of the Upper Paleolithic. Considering it "evidence of the modern human mind at work," Leakey (1984:93-94) describes this period as "unlike previous eras, when stasis dominated, ... [with] change being measured in millennia rather than hundreds of millennia." This period exhibits more innovation than in the previous six million years of human evolution (Mithen, 1998). It marks the beginnings of traits considered diagnostic of behavioral modernity, including a more organized, strategic, season-specific style of hunting involving specific animals at specific sites, elaborate burial sites indicative of ritual and religion, evidence of dance, magic, and totemism, colonization of Australia, and replacement of Levallois tool technology by blade cores in the Near East. In Europe, complex hearths and art appeared, including cave paintings of animals, decorated tools and pottery, bone and antler tools with engraved designs, ivory statues of animals and sea shells, and personal decoration such as beads, pendants, and perforated animal teeth, many of which may have indicated social status.

Whether this period was a genuine revolution culminating in behavioral modernity is hotly debated because claims to this effect are based on the European Paleolithic record, and largely exclude the African record (Fisher & Ridley, 2013). Many artifacts associated with a rapid transition to behavioral modernity 40-50,000 years ago in Europe are found in the African Middle Stone Age tens of thousands of years earlier, which pushes the cultural transition more closely into alignment with the transition to anatomical modernity and the associated changes to FOXP2. What is clear is that modern behavior appeared in Africa between 100,000 to 50,000 years ago, and spread, resulting in displacement of the Neanderthals in Europe (Fisher & Ridley, 2013). From this point on there was only one hominid species: modern *Homo sapiens*. Despite a lack of overall increase in cranial capacity, the prefrontal cortex, and more particularly the orbitofrontal region, increased significantly in size, in what was most likely a time of major neural reorganization (Morgan, 2013).

Thus, a period of significant FOXP2 evolution overlaps with both the transition to anatomical modernity and the cultural transition of the Middle-Upper Paleolithic. This suggests that that FOXP2 may have played a role in the onset of modern human cognitive abilities.

## Contextual Focus as Explanation for the Creative Explosion

We have seen evidence that FOXP2 plays a role in the evolution of complex cognitive abilities including language but this relationship is not solely mediated through its effects on the capacity for imitation. We now provide the necessary background to consider an alternative scenario.

It has been proposed that the Paleolithic transition reflects fine-tuning of the capacity to subconsciously shift between different modes of thought (Gabora, 2003 Gabora & Kaufman, 2010; Gabora & Kitto, 2013). The ability to shift between different modes was referred to as *contextual focus* (CF) because it requires the ability to focus or defocus attention in response to the context or situation one is in. CF is in some ways similar to the notion of *dual processing*, according to which we use *explicit* cognition for conscious analysis of the task at hand, and *implicit* cognition for free association and quick 'gut' responses (Evans, 2008; Frankish, 2011; Nosek, 2007; Willingham & Goedert-Eschmann, 1999). The creativity literature makes a similar distinction between (1) divergent or *associative* processes during idea generation, while (2) convergent or *analytic* processes predominate during the refinement, implementation, and testing of an idea (Finke, Ward, & Smith, 1992). Associative thought maps onto implicit cognition, while analytic thought maps onto explicit cognition (see Sowden, Pringle, & Gabora, in press, for a comparison and discussion of the relationship dual processing theory and dual theories of creativity).

The proposal then is that the onset of CF in the Paleolithic was made possible through fine-tuning of the biochemical mechanisms underlying the capacity to shift between these modes, depending on the situation, by varying the specificity of the activated memory region (Gabora, 2003). Defocused attention, by diffusely activating a wide region of memory, is conducive to associative thought; it enables obscure (but potentially relevant) aspects of the situation to come into play. It is useful for the generation of new connections when one is stuck in a rut. Focused attention is conducive to analytic thought; memory activation is constrained enough to hone in and perform logical mental operations on the most clearly relevant aspects. It is reasonable that one or more genes would have been involved in the fine-tuning of the biochemical mechanisms underlying the capacity to subconsciously shift between these two processing modes. Regions consisting of many cell assemblies participate in the encoding of more memories than regions containing few.

## Formal Models of Contextual Focus

The hypothesis that CF is responsible for the burst of innovation in the Paleolithic is supported by formal models of CF. A mathematical model of CF was developed (Gabora & Aerts, 2009), and a version of this model was consistent with experimental data from a study in which participants were asked to rate the typicality of exemplars of a concept for different contexts (Veloz, Gabora, Eyjolfson, & Aerts, 2011). Introducing measures of state robustness and context relevance, the modeled showed that a shift to a more associative processing mode (by varying the exemplar typicality threshold) is accompanied by changes in the relevance of different contexts, such that seemingly atypical states of the concept can become typical states. This result is important because viewing a familiar concept from a new context and thereby making new connections is the essence of creativity; it underlies not just the generativity of language but the characteristically human ability to adapt old ways to new circumstances.

The effect of CF at the level of society was investigated using a computational model of cultural evolution consisting of an artificial society of neural-network based agents that invent and imitate ideas for cultural outputs. CF was introduced into an artificial society by giving agents the capacity to shift between different processing modes depending on the effectiveness of their current cultural output (Gabora & Firouzi, 2012). The fitness of cultural outputs was significantly higher with CF than without it, as shown in Figure 1. The diversity, or number of cultural outputs, showed the typical pattern of an increase as the space of possibilities is explored followed by a decrease as agents converge on the fittest outputs. However, with CF, this occurs more quickly, and although early in a run diversity is lower with CF (because agents more quickly hone in on superior cultural outputs) in the long run diversity is higher, as shown in Figure 2 (because a greater number of superior cultural outputs are found). Together, Figures 1 and 2 show that the onset of CF provides a computationally feasible explanation for how the increased utility and variety of artifacts in the Middle/Upper Paleolithic came about.

## The Neurological Basis of Contextual Focus

To see why the onset of CF could reasonably be attributed to fine-tuning at the neurological level and give rise to a Paleolithic burst of creativity we need to examine the features of associative memory that allow creative connections to be made. The notion that diffuse activation is conducive to associative thought while activation of a narrow receptive field is conducive to analytic thought is consistent with the architecture of associative memory (Gabora, 2000; Gabora & Ranjan, 2012). Memories are encoded in neurons that are sensitive to ranges (or values) of *microfeatures* (Churchland & Sejnowski, 1992). Each neuron responds maximally to a particular microfeature and responds to a lesser extent to related microfeatures, an organization referred to as *coarse coding* (Hubel & Wiesel, 1965). An item in memory is *distributed* across cell assemblies that contain many neurons; each neuron participates in the storage of many items (Hinton, McClelland, & Rumelhart, 1986). Memory is also *content addressable:* there is a systematic relationship between the content of a representation and the neurons where it gets encoded. Thus representations that get encoded in overlapping regions of memory share features, and representations can be evoked by stimuli that are similar or "resonant" in some (perhaps context-specific) way (Hinton,

McClelland, & Rumelhart, 1986; Marr, 1969).

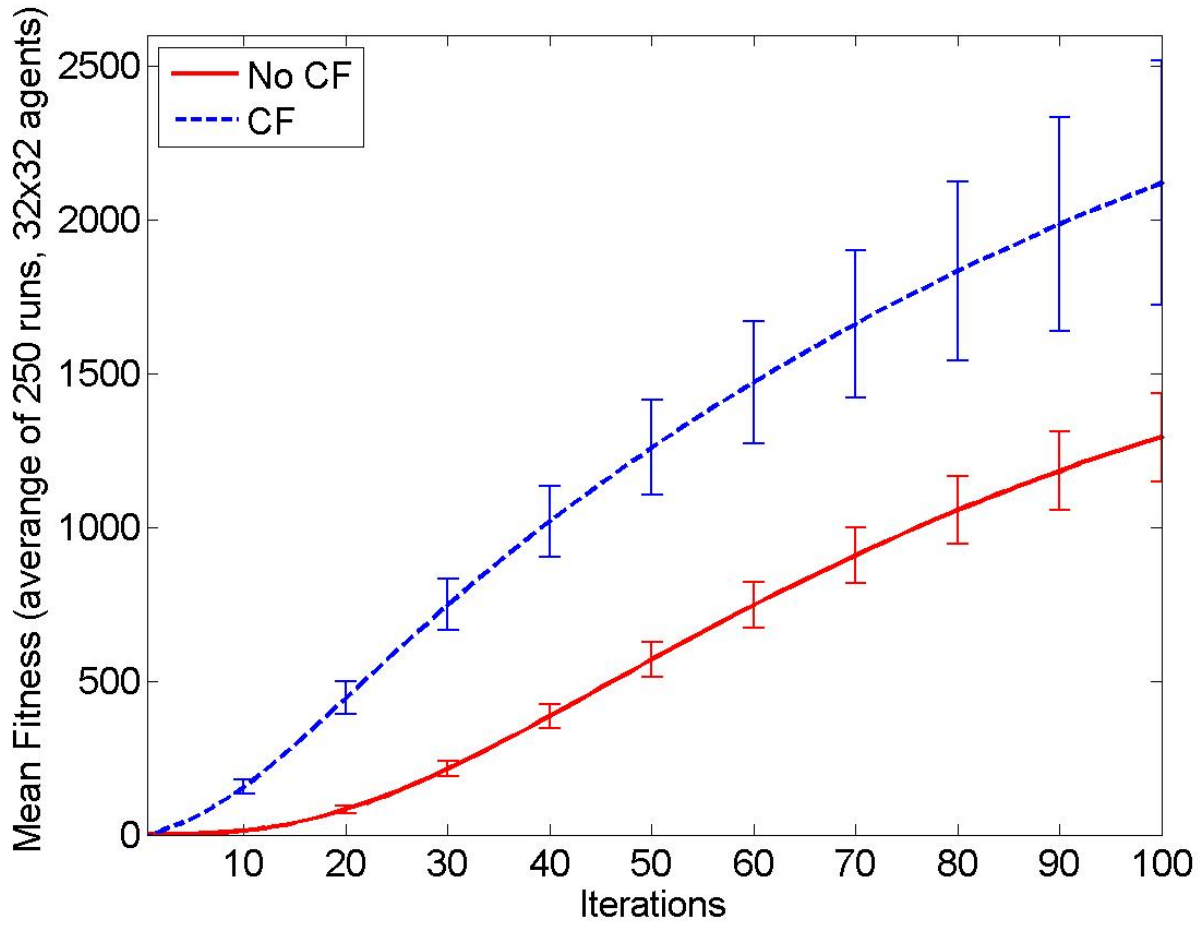

Figure 1. The mean fitness of cultural outputs generated by agents in a computational model of cultural evolution was higher in agents with CF (those with the ability to shift between different processing modes).

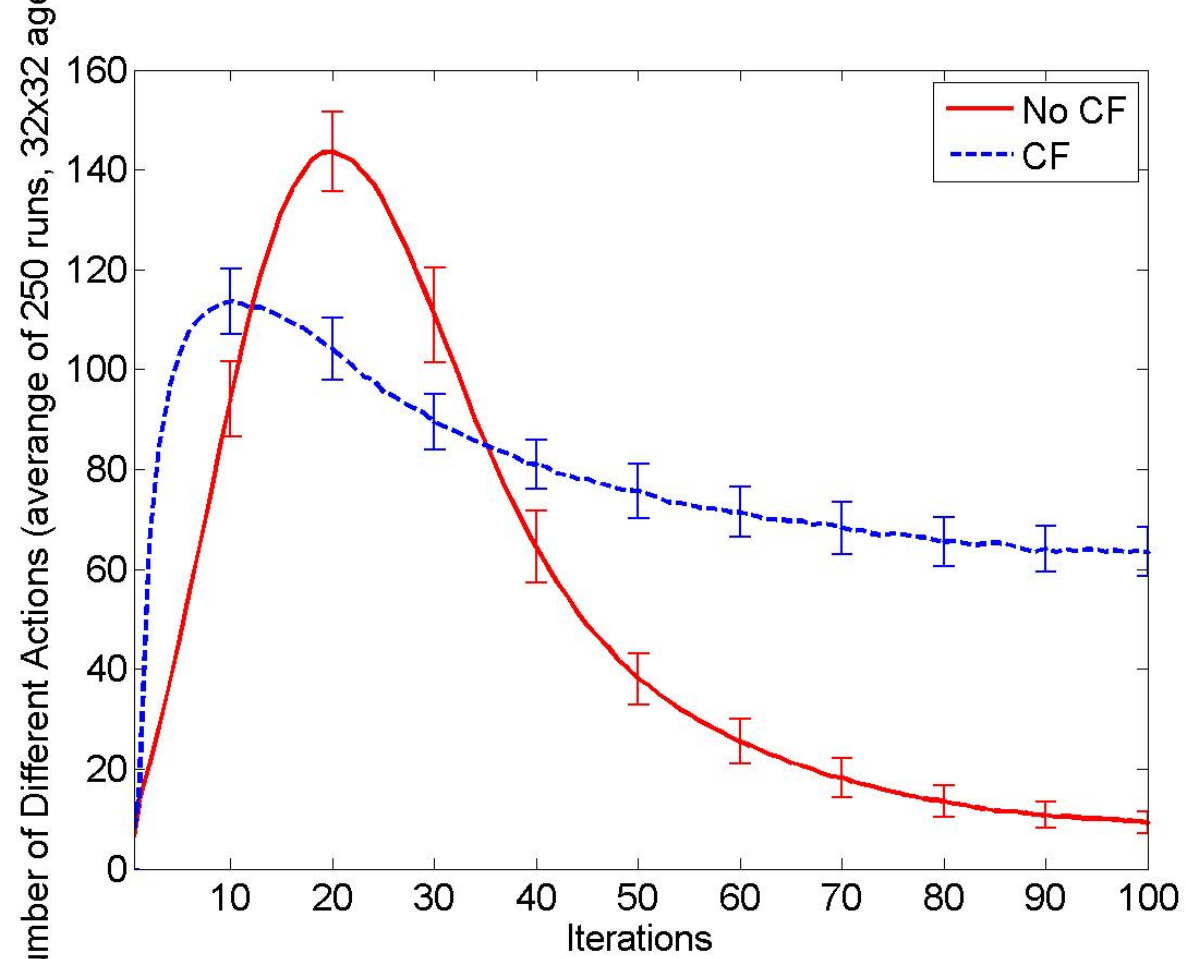

Figure 2. Figure 1. The diversity of cultural outputs generated by agents with CF increases more quickly and stabilizes at a higher value than for agents without CF.

These properties of associative memory—distributed representation, coarse coding, and content addressability—make possible both the capacity to stay on task during analytic thought, and the forging of unusual yet relevant connections during associative thought (Gabora, 2000; Gabora & Ranjan, 2012). Each thought may activate more or fewer cell assemblies depending on the nature of the task at hand. Focused attention is conducive to analytic thought because memory activation is constrained enough to zero-in and specifically operate on the most defining properties. Defocused attention, by diffusely activating a diversity of memory locations, is conducive to associative thought; obscure (but potentially relevant) properties of the situation come into play. Thus while in an analytic mode of thought the concept TOOL might only activate a hand axe, the most commonly used tool, in an associative mode of thought, all sorts of items in one's environment might be viewed as a potential tool depending on what one wants to accomplish. Once it was possible to shift between these modes of thought, cognitive processes requiring either analytic thought, associative thought or both could be carried out more effectively, and indeed the fruits of one mode of thought could become ingredients for the other mode, thereby facilitating the forging of a richly integrated internal model of the world. Hence we propose that the Paleolithic transition reflects a mutation to FOXP2 that caused fine-tuning of the capacity to spontaneously shift between associative and analytic modes depending on the situation by varying the specificity of the activated memory region.

## The Evolutionary Origins of Contextual Focus

We now present our argument for how FOXP2 made possible the explosion of creativity and onset of language in the Middle / Upper Paleolithic. Amino acid substitutions that differentiate the human genome FOXP2 gene from that of chimpanzees led to the appearance of anatomically modern humans. This genetic modification changed the functioning of neurons in the basal ganglia that contribute to cognitive flexibility. These neurons show greater synaptic plasticity and dendrite length in humans compared to chimpanzees. These changes enhanced the efficiency of neural cortico-basal ganglia circuits in the human brain. It enabled individuals to modulate the degree of cognitive flexibility by spontaneously adjusting how much detail of any given thought or experience impact associative memory. Humans became able to shift between focused and defocused modes of attention, and thereby engage in CF. By tuning one's mode of thought in response to the needs of the present moment, CF allowed information to be processed at different degrees of granularity, and from different perspectives.

It is thus by way of CF that hominids became able to combine words into an infinite variety of sentences, to spontaneously change the direction of abstract thought, and to adapt behavior to changing circumstances. Individuals could engage in analytic thought as a default, but switch to associative thought when they were stuck, or when they wanted to express themselves or produce outputs that were aesthetically satisfying. Associative thought enabled them to connect ideas in new ways, resulting in creative ideas. These new mental combinations resulted in advanced tools, elaborate burials, and many forms of art and jewelry.

### Creativity, Evolution, ADHD, and Schizophrenia

Further support for the proposal that FOXP2 played a role in the creative explosion of the Paleolithic and the onset of complex language comes from the literature on schizophrenia and attention deficit hyperactivity disorder (ADHD). Attention-deficit/hyperactivity disorder (ADHD) is a common psychiatric disorder with symptoms of inattention, hyperactivity, and/or impulsivity that occurs in approximately 4–5% of children and persists into adulthood in at least 30% of patients diagnosed during infancy (Ribases et al., 2012). Genes that contribute to the acquisition of reading and spelling skills and play a role in speech and language are strong candidates to be involved in both ADHD and learning problems. FOXP2 affects activity of the corticobasal ganglia circuits known to contribute to ADHD (Ribases et al., 2012).

Schizophrenia is a serious mental disorder characterized by delusions, hallucinations, disorganized speech, catatonia, paranoia, keen interest in religion, emotional flattening, and creativity. Hallucinations may take the form of verbal running commentary, or multiple voices conversing with each other. There is evidence that neural connections that predispose one to schizophrenia contribute to creativity (Cropley & Sikand, 1973). Genes that affect the dorsolateral prefrontal and orbitofrontal cortices differently individuals with schizophrenia exhibit positive selection for differential expression between humans and chimpanzees (Morgan, 2013). Moreover, a polymorphism in FOXP2 can cause language deficits in patients with schizophrenia, suggesting that increased risk of schizophrenia is the price *Homo sapiens* pay for language and creativity (Sanjuan, Tolosa, Dagnall, Molto, & de Frutos, 2010). Moreover, schizophrenia is sometimes associated with both language impairment and above-average creativity (Kuttner Lorincz, & Swan, 1967). These findings suggest that the creative explosion of the Middle-Upper Paleolithic including the onset of complex language may have indirectly selected for genes involved in schizophrenia.

## Discussion

The overall thread of our argument is synthesized in Figure 3. We propose that a form of FOXP2 that evolved during the Paleolithic was responsible for the neurological underpinnings of the ability to shift between associative and analytic modes of thought. This enabled hominids to process information at multiple levels of detail and from different perspectives. It generated a need to think through and communicate complex thoughts. This brought selective pressure for the capacity to express oneself, thereby creating an environment conducive to the emergence of complex language including recursion, grammar, word inflections, syntactical structure, and comprehension.

The proposal is consistent with findings that FOXP2 is associated with cognitive abilities that do not involve language, and with findings that non-language creative abilities arose at approximately the same time as complex language. The argument is also consistent with the view that what is at the core of our uniquely human cognitive abilities is the capacity to place things in context, or see things from different perspectives (Gabora & Kaufman, 2010). This enables us to put our own spin on the inventions of others, modifying them to suit our own needs and tastes, leading to new innovations that build cumulatively on previous ones. It permits us to modify thoughts, impressions, and attitudes by thinking about them in the context of each other, and weaving them into an integrated structure that defines who we are in relation to the world. Our compunction to put our own spin on the ideas and inventions of others results in accumulative cultural change, referred to as the *ratchet effect* (Tomasello, Kruger, & Ratner, 1993).

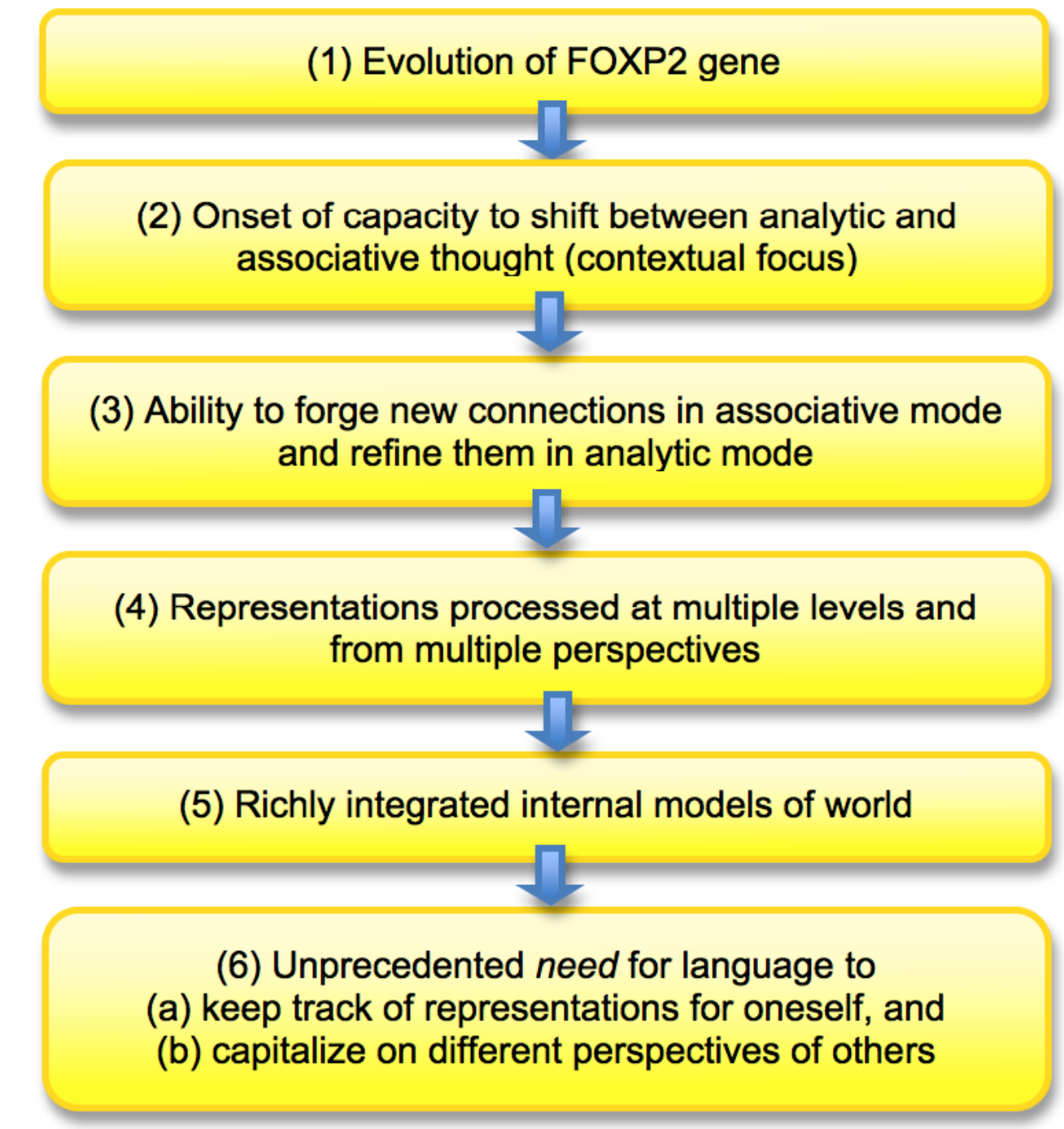

Figure 1: Schematic illustration of the proposed relationship between FOXP2, language, and contextual focus.

The picture presented here is speculative and incomplete. Likely other genes play a role in the capacity for CF, and other cognitive mechanisms are involved. Future research will aim at developing means of testing the hypothesis, and elucidating the neural mechanisms by which FOXP2 could enable a shift between different modes of thought during different phases of problem solving and other creative tasks.

## Acknowledgments

We are grateful for funding to the second author from the Natural Sciences and Engineering Research Council of Canada and the Flemish Fund for Scientific Research, Belgium, and to Simon Tseng who made the graphs.